\documentclass[twocolumn,prb,amsmath,amssymb,floatfix]{revtex4}

\usepackage{bbold}
\usepackage{bm}
\usepackage{color}
\usepackage{dcolumn}
\usepackage{feynmf} 
\usepackage{graphics}
\usepackage{graphicx}
\usepackage{subfigure}
\usepackage{slashed}
\usepackage{tensor}
\usepackage{placeins}
\usepackage{natbib}

\def\al{\alpha}

\def\veps{\varepsilon}

\def\be{\begin{equation}}
\def\ee{\end{equation}}
\def\bea{\begin{eqnarray}}
\def\eea{\end{eqnarray}}
\def\bse{\begin{subequations}}
\def\ese{\end{subequations}}
\def\bc{\begin{center}}
\def\ec{\end{center}}

\def\ra{\rightarrow}

\def\nonum{\nonumber}

\def\D{{\rm d}}

\newcommand{\comment}[1]{}

\catcode`,\active

\catcode`\,12

\begin{document}

\title{Two-loop fermion self-energy and propagator in reduced QED$_{3,2}$}

\author{S.~Teber$^{1,2}$}
\affiliation{
$^1$Sorbonne Universit\'es, UPMC Univ Paris 06, UMR 7589, LPTHE, F-75005, Paris, France.\\
$^2$CNRS, UMR 7589, LPTHE, F-75005, Paris, France.}

\date{\today}

\begin{abstract}
We compute the two-loop fermion self-energy in massless reduced quantum electrodynamics (RQED) for an arbitrary gauge in the case where
the photon field is three-dimensional and the fermion field two-dimensional: super-renormalizable RQED$_{3,2}$ with $N_F$ fermions.
We find that the theory is infrared finite at two-loop and that finite corrections to the fermion propagator have a remarkably simple form.
\end{abstract}

\maketitle

\begin{fmffile}{fmfsigma32}

%
\begin{figure}[tl]
   \includegraphics{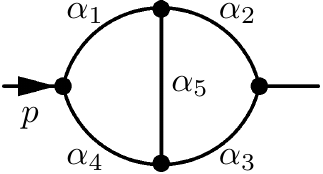}
   \caption{\label{fig:J}
   Two-loop massless propagator diagram.}
\end{figure}
%

One of the building blocks of multi-loop calculations is the two-loop massless propagator diagram, see Fig.~\ref{fig:J}:
\begin{widetext}
\bea
\int \int \frac{\D^{d_e} k_1 \, \D^{d_e} k_2}{[-(k_1+p)^2]^{\al_1}\,[-(k_2+p)^2]^{\al_2}\,[-k_2^2]^{\al_3}\,[-k_1^2]^{\al_4}\,[-(k_2-k_1)^2]^{\al_5}}
= - \frac{\pi^{d_e}}{(-p^2)^{\sum_{i=1}^5 \al_i - d_e}}\,G(\al_1,\al_2,\al_3,\al_4,\al_5)\, ,
\label{def:2loopmassless}
\eea
\end{widetext}
where $G$ is the so-called coefficient function of the diagram, $\al_i$ are arbitrary indices 
and $p$ is an external momentum in a Minkowski space-time of dimensionality $d_e$.
This diagram is at the heart of numerous radiative correction calculations in quantum field theory
and associated to the development of sophisticated methods such as, {\it e.g.}, 
the Gegenbauer polynomial technique~\cite{ChetyrkinKT80,Kotikov96}, integration by parts~\cite{VasilievPK81,ChetyrkinT81}, and
the method of uniqueness~\cite{VasilievPK81,Usyukina83,Kazakov84}, see Ref.~[\onlinecite{Grozin-lecture12}] for a historical review on this diagram.
In the case where all indices are integers, this diagram is well known and can be expressed in terms of recursively one-loop diagrams.  
When all indices are arbitrary, the result is highly non-trivial and can be represented~\cite{Bierenbaum} as a combination
of two-fold series. In some intermediate cases, simpler forms can be obtained.~\cite{VasilievPK81,Kazakov85,KivelSV93+94,Kotikov96,BroadhurstGK97,BroadhurstK96,KotikovT13a}
In particular, in Ref.~[\onlinecite{Kotikov96}], an ingenious transformation was found from Gegenbauer two-fold series
to one-fold ${}_3F_2$-hypergeometric series of unit argument for a complicated class of diagrams having two integer indices on adjacent lines and three other arbitrary indices.
For this class of diagrams, similar results have been found in Ref.~[\onlinecite{BroadhurstGK97}] using an ansatz to solve the
recurrence relations arising from integration by parts. In Ref.~[\onlinecite{Kotikov96}], the results were applied to the computation
of a diagram with a single non-integer index on the central line. This important diagram appears in various calculations, see, {\it e.g.}, 
Refs.~[\onlinecite{VasilievPK81,Kazakov85,KivelSV93+94,Teber12,KotikovT13a}]; it was shown 
in Ref.~[\onlinecite{Kotikov96}] to reduce to a single ${}_3F_2$-hypergeometric series of unit argument.
More recently, in Ref.~[\onlinecite{KotikovT13b}], the results of [\onlinecite{Kotikov96}] were applied to the case involving two arbitrary indices on non adjacent lines.
In this case, the corresponding coefficient function:
\be
G(\al,1,\beta,1,1) =
C_D\left[ \quad \parbox{16mm}{
    \begin{fmfgraph*}(16,14)
      \fmfleft{i}
      \fmfright{o}
      \fmfleft{ve}
      \fmfright{vo}
      \fmftop{vn}
      \fmftop{vs}
      \fmffreeze
      \fmfforce{(-0.1w,0.5h)}{i}
      \fmfforce{(1.1w,0.5h)}{o}
      \fmfforce{(0w,0.5h)}{ve}
      \fmfforce{(1.0w,0.5h)}{vo}
      \fmfforce{(.5w,0.95h)}{vn}
      \fmfforce{(.5w,0.05h)}{vs}
      \fmffreeze
      \fmf{plain}{i,ve}
      \fmf{plain,left=0.8}{ve,vo}
      \fmf{phantom,left=0.5,label=$\al$,l.d=-0.01w}{ve,vn}
      \fmf{phantom,right=0.5,label=$1$,l.d=-0.01w}{vo,vn}
      \fmf{plain,left=0.8}{vo,ve}
      \fmf{phantom,left=0.5,label=$\beta$,l.d=-0.01w}{vo,vs}
      \fmf{phantom,right=0.5,label=$1$,l.d=-0.01w}{ve,vs}
      \fmf{plain,label=$1$,l.d=0.05w}{vs,vn}
      \fmf{plain}{vo,o}
      \fmffreeze
      \fmfdot{ve,vn,vo,vs}
    \end{fmfgraph*}
}
\quad \right]\, ,
\label{G}
\ee
was shown to reduce to two ${}_3F_2$-hypergeometric series of argument $1$. 

\begin{figure}
\includegraphics{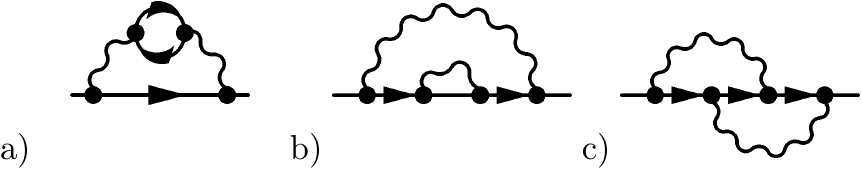}
\caption{\label{fig:s}
   Two-loop fermion self-energy diagrams.}
\end{figure}

In Ref.~[\onlinecite{KotikovT13b}], Eq.~(\ref{G}) appeared in the computation of the two-loop fermion self-energy in reduced quantum electrodynamics (RQED), [\onlinecite{GorbarGM01}], or
RQED$_{d_\gamma,d_e}$, see also Refs.~[\onlinecite{Marino93+DoreyM92+KovnerR90}] in relation with RQED$_{4,3}$. 
In the general case, this relativistic model describes the interaction of an abelian $U(1)$ gauge field
living in $d_\gamma$ space-time dimensions with a fermion field localized in a reduced space-time of $d_e$ dimensions ($d_e \leqslant d_\gamma$).
In RQED$_{d_\gamma,d_e}$, while the bubble and rainbow diagrams, Figs.~\ref{fig:s}\,a) and \ref{fig:s}\,b), respectively,
naturally reduce to recursively one-loop diagrams, the crossed photon diagram, Fig.~\ref{fig:s}\,c), involves a
contribution of the type Eq.~(\ref{G}) with the indices given by:
\be
\al = \beta = 1- \veps_e\, ,
\ee
where, following the notation of Ref.~[\onlinecite{KotikovT13b}], $d_\gamma = 4 -2\veps_\gamma$ and $d_e = 4 -2 \veps_e - 2 \veps_\gamma$.
In the case of usual QEDs, {\it e.g.}, QED$_4$ and QED$_3$, the parameter $\veps_e=0$ and all indices are integers.
Reduced models appear to be more complicated, {\it a priori}, as they generally involve non-integer indices, {\it e.g.},
$\veps_e = 1/2$ and $\veps_\gamma \ra 0$ for RQED$_{4,3}$ which corresponds to the ultrarelativistic limit of an undoped graphene monolayer.
This complication turned out to be overcome, in the case of RQED$_{4,d_e}$, by the presence of a coefficient $\veps_\gamma \ra 0$ in factor of
the ultra-violet (UV) convergent Eq.~(\ref{G}) in the expression of the self-energy.

In this Brief Report we complete the previous study by examining the case of RQED$_{3,2}$ which is interesting from the field theory point of view as it
does require the computation of Eq.~(\ref{G}) for $\veps_e = 1/2$ and
$\delta_\gamma = \veps_\gamma - 1/2 \ra 0$. We shall show that the formulas of Ref.~[\onlinecite{KotikovT13b}] are extremely convenient to perform such a task.
Moreover, similarly to QED$_3$ ($\veps_e = 0$ and $\delta_\gamma \ra 0$), RQED$_{3,2}$ is super-renormalizable and therefore asymptotically free.
However, contrary to QED$_3$ where infra-red (IR) divergences yield an anomalous dimension to the fermion field at two loop, RQED$_{3,2}$
is finite at two-loop and the corrections to the fermion propagator take a very simple form, as will be shown below.

\begin{widetext}
Following [\onlinecite{KotikovT13b}], we start by considering the fermion self-energy up to two loops in RQED$_{3,d_e}$:
\bea
\Sigma_{V}(p^2) = \frac{\tilde{\al}}{4\pi}\,e^{(\gamma_E-L_p)\delta_\gamma}\,\sigma_1(\varepsilon_e,\delta_\gamma,a)
+ \left( \frac{\tilde{\al}}{4\pi} \right)^2\, e^{2(\gamma_E-L_p)\delta_\gamma}\,\sigma_2(\varepsilon_e,\delta_\gamma,a)\, ,
\label{Ren:def:sigma2} 
\eea
where $L_p = \ln(-p^2/\mu^2)$, $\tilde{\al}$ is a momentum-dependent dimensionless coupling constant ($e^2$ has dimension of mass in RQED$_{3,d_e}$) defined as:
$\tilde{\al} = e^2 / \sqrt{-4\pi p^2}$
and we have used the fact that charge does not renormalize in RQED$_{3,d_e}$. The one-loop contribution reads:
\bea
\sigma_1 = \Gamma(1-\varepsilon_e)\,\frac{1-2\veps_e-2\delta_\gamma}{2}\,\left(\frac{\varepsilon_e}{1-\veps_e-2\delta_\gamma}-a \right)\,G(1,1-\varepsilon_e)\, ,
\label{sigma1-sep} 
\eea
where $a$ is a gauge fixing parameter ($a=1$ in the Feynman gauge) that we shall keep arbitrary in what follows 
and $G(\al,\beta)$ is the coefficient function of the one-loop massless propagator diagram:
\be
G(\al,\beta) = \frac{a(\al)a(\beta)}{a(\al+\beta-d_e/2)}\, \qquad a(\al) = \frac{\Gamma(d_e/2-\al)}{\Gamma(\al)}\, .
\ee
The two-loop function corresponds to the sum of the three diagrams in Fig.~\ref{fig:s}, $\sigma_2 = \sigma^{(2)}_a+\sigma^{(2)}_b+\sigma^{(2)}_c$,
where the first two diagrams yield ($N_F$ is the number of massless fermion fields)
\begin{subequations}
\label{ren-sigma2s-a+b}
\bea
&&\sigma^{(2)}_a
= -2N_F\,\Gamma^2(1-\varepsilon_e)\,\frac{(1-2\varepsilon_e-2\delta_\gamma)^2}{3-2\varepsilon_e+2\delta_\gamma}\,G(1,1)G(1,1/2-\varepsilon_e+\delta_\gamma)\, ,
\\
&&\sigma^{(2)}_b =
2\,\Gamma^2(1-\varepsilon_e)\,\frac{\delta_\gamma(1-2\varepsilon_e-2\delta_\gamma)\left[\varepsilon_e-a(1-\varepsilon_e-2\delta_\gamma)\right]^2}
{(1+2\delta_\gamma) (1-\varepsilon_e-2\delta_\gamma)}\,G(1,1-\varepsilon_e)G(1-\varepsilon_e, 1/2 + \delta_\gamma )\, ,
\eea
\end{subequations}
and the third diagram can be further separated into three parts $\sigma^{(2)}_c = \sigma^{(2)}_{c_1} + \sigma^{(2)}_{c_2} + \sigma^{(2)}_{c_3}$
where
\begin{subequations}
\label{ren-sigma2s-c}
\bea
&&\hskip -12pt \sigma^{(2)}_{c_1} = \Gamma^2(1-\varepsilon_e)\, \frac{1-2\veps_e-2\delta_\gamma}{2}\,G^2(1,1-\veps_e)\, 
\left [ 1 + 2\veps_e + 2\delta_\gamma +(1-a)\frac{(1-2\veps_e-2\delta_\gamma)^2}{1-\veps_e-2\delta_\gamma}
-\frac{(1-a)^2}{2}\,(1-2\veps_e-2\delta_\gamma) \right .
 \\
&&\left . + 2\,\frac{(\veps_e+2\delta_\gamma)\,[8-(3-2\delta_\gamma)(1+2\veps_e+2\delta_\gamma)]}{(1+2\veps_e+6\delta_\gamma)(1-2\veps_e-6\delta_\gamma)} +
\frac{4\veps_e}{1-\veps_e-2\delta_\gamma} 
 - \frac{2\veps_e}{1-2\veps_e-6\delta_\gamma}\,\left(5+2\veps_e+2\delta_\gamma-4\,\frac{\veps_e+2\delta_\gamma}{1-\veps_e-2\delta_\gamma} \right) \right]\, ,
\nonum \\
&&\hskip -12pt \sigma^{(2)}_{c_2} = -\Gamma^2(1-\varepsilon_e)\, \frac{1-2\veps_e-2\delta_\gamma}{2}\,G(1,1-\veps_e)\,G (1-\varepsilon_e, 1/2 + \delta_\gamma )\,
\left[ 2 - 4\veps_e - 4\delta_\gamma + 4\,\frac{1-2\veps_e-2\delta_\gamma}{1-\veps_e-2\delta_\gamma}-16\,\frac{1-\delta_\gamma}{1+2\delta_\gamma} \right . 
\\
&&\left . -4(1-a)\,\frac{\delta_\gamma (1-2\veps_e-2\delta_\gamma)}{1-\veps_e-2\delta_\gamma}  + 2(1-a)^2\,\delta_\gamma +
2\,\frac{\veps_e(5+2\veps_e+2\delta_\gamma)}{\veps_e+2\delta_\gamma}+4\,\frac{\veps_e(1-2\delta_\gamma)}{(1+2\delta_\gamma)(1-\veps_e-2\delta_\gamma)}\right]\, ,
\nonum \\
&&\hskip -12pt \sigma^{(2)}_{c_3} = \Gamma^2(1-\varepsilon_e)\, \frac{1-2\veps_e-2\delta_\gamma}{2}\, G(1-\veps_e,1,1-\veps_e,1,1)\,
\frac{(1+2\delta_\gamma)\,[8 - (3-2\delta_\gamma)(1+2\veps_e+2\delta_\gamma)]}{(1-2\veps_e-6\delta_\gamma)(1+2\veps_e+6\delta_\gamma)}\, .
\eea
\end{subequations}
Straightforward application of the above equations to RQED$_{3,2}$ ($\veps_e = 1/2$ and $\delta_\gamma \ra 0$) yields the following expansions:
\begin{subequations}
\bea
&& e^{(\gamma_E-L_p)\delta_\gamma}\,\sigma_1 = \sqrt{\pi}\,(1-a) + \sqrt{\pi}\,\delta_\gamma\,[4-(1-a)\bar{L}_p] + O(\delta_\gamma^2)\, ,
\\
&& e^{2(\gamma_E-L_p)\delta_\gamma}\,\sigma_a^{(2)} = -4 \pi N_F + 4\pi N_F\,\delta_\gamma\,(1+2 \bar{L}_p - 8 \ln 2 ) + O(\delta_\gamma^2)\, ,
\\
&& e^{2(\gamma_E-L_p)\delta_\gamma}\,\sigma_b^{(2)} = \pi (1-a)^2 + 2\pi (1-a)\,\delta_\gamma\,[1+3a -(1-a)(\bar{L}_p-6 \ln 2)]  + O(\delta_\gamma^2)\, ,
\eea
\end{subequations}
%
%
\begin{subequations}
\bea
&& e^{2(\gamma_E-L_p)\delta_\gamma}\,\sigma_{c_1}^{(2)} = -\frac{\pi}{6\delta_\gamma^2} - \frac{\pi\,\big(3 - \bar{L}_p \big)}{3 \delta_\gamma}
+\pi\,\bigg(3-(1-a)^2 + \frac{5}{6}\,\zeta_2 + 2 \bar{L}_p - \frac{1}{3}\,\bar{L}_p^2 \bigg) 
\nonum \\
&& + \pi\,\delta_\gamma\,\bigg( \frac{47}{3} + 5 \zeta_2 + \frac{55}{9}\,\zeta_3 -4 \bar{L}_p - \frac{5}{3}\,\zeta_2\,\bar{L}_p -2 \bar{L}_p^2 +\frac{2}{9}\,\bar{L}_p^3
+8a + 2 a^2 \bar{L}_p-4a\bar{L}_p \bigg) + O(\delta_\gamma^2)\, ,
\\
&& e^{2(\gamma_E-L_p)\delta_\gamma}\,\sigma_{c_2}^{(2)} = \pi\,\big(8 - (1-a)^2 \big ) + 2\pi \,\delta_\gamma\,\bigg(4a -16 - (8-(1-a)^2)(\bar{L}_p - 6\ln 2) \bigg) + O(\delta_\gamma^2)\, ,
\\
&& e^{2(\gamma_E-L_p)\delta_\gamma}\,\sigma_{c_3}^{(2)} = \frac{\pi}{6}\,\frac{(1+2\delta_\gamma)\,[4-(3-2\delta_\gamma)(1+\delta_\gamma)]}{1+3\delta_\gamma}\,
e^{2(\gamma_E-L_p)\delta_\gamma}\,G(1/2,1,1/2,1,1)\, ,
\label{sigma-c3}
\eea
\end{subequations}
where $\bar{L}_p = L_p + 4 \ln 2$, $\sigma_{c_1}^{(2)}$ is explicitly IR singular and the last term contains a contribution from the 
complicated diagram, Eq.~(\ref{G}), which cannot be reduced to products of one-loop massless functions.

Among the various forms derived for $G(\al,1,\beta,1,1)$ in Ref.~[\onlinecite{KotikovT13b}], the most convenient one for the present application
is Eq.~(B10) in that paper. Together with (B11) and in the case of RQED$_{3,2}$, these equations yield:
\begin{subequations}
\bea
&&G(1/2,1,1/2,1,1) = -24\,\frac{\delta_\gamma\,(1+3\delta_\gamma)}{1+2\delta_\gamma}\,
\frac{\Gamma^2(1/2-\delta_\gamma) \Gamma(1-\delta_\gamma) \Gamma(1+2\delta_\gamma)}{\Gamma(1/2) \Gamma(1-2\delta_\gamma)\Gamma(1-3\delta_\gamma)}\,I(1/2)\, ,
\\
&& I(1/2) = \frac{\Gamma(1/2)}{\Gamma(2+2\delta_\gamma)}\,\frac{\pi \, \sin[\pi \delta_\gamma ]}{\sin[\pi(1/2+2\delta_\gamma)]\sin[\pi(1/2-\delta_\gamma)]}
 + \sum_{n=0}^{\infty} \frac{\Gamma(n-2\delta_\gamma)\Gamma(n+1)}{n!\,\Gamma(n+3/2)}\,\frac{1}{n-1/2-2\delta_\gamma}
\label{I} \\
&& + \frac{1+4\delta_\gamma}{4\delta_\gamma}\,\frac{\Gamma(1/2+\delta_\gamma) \Gamma(1-\delta_\gamma)}{\Gamma(1/2-2\delta_\gamma)\Gamma(1+2\delta_\gamma)}\,
\frac{\sin[\pi(1/2+2\delta_\gamma)]}{\sin[\pi(1/2-\delta_\gamma)]}\,
\sum_{n=0}^{\infty} \frac{\Gamma(n-2\delta_\gamma)\Gamma(n-1-3\delta_\gamma)}{n!\,\Gamma(n-1/2-3\delta_\gamma)}\,\frac{1}{n-1/2-2\delta_\gamma}\, .
\nonum
\eea
\end{subequations}
Indeed, under this form, the $\delta_\gamma$-expansion of the hypergeometric functions with non-integer parameters is most easily done. We shall carry such expansion
up to $O(1)$ which is what is needed for $I$ in order to expand the $G$-function up to $O(\delta_\gamma)$.
The first term in Eq.~(\ref{I}) is of $O(\delta_\gamma)$ and can be neglected. 
The second term is singular and expands as:
\bea
\sum_{n=0}^{\infty} \frac{\Gamma(n-2\delta_\gamma)\Gamma(n+1)}{n!\,\Gamma(n+3/2)}\,\frac{1}{n-1/2-2\delta_\gamma}
= \frac{\Gamma(1-2\delta_\gamma)}{\Gamma(3/2)}\, \bigg[ \frac{1}{\delta_\gamma} + 4 G - 6 + O(\delta_\gamma) \bigg]\, ,
\label{term2}
\eea
where $G$ is Catalan's constant. The third term is conveniently split into two parts following the property that 
\be
\frac{1}{(n-1-3\delta_\gamma)} \,
\frac{(n-1/2-3\delta_\gamma)}{(n-1/2-2\delta_\gamma)} = 
\frac{1}{(1+2\delta_\gamma)} \left[\frac{1}{(n-1-3\delta_\gamma)}
+ \frac{2\delta_\gamma}{n-1/2-2\delta_\gamma} \right] \, .
\label{split}
\ee
The first term in the r.h.s of Eq.~(\ref{split}) can be summed exactly as a ${}_2F_1$-series of unit argument and is singular:
\bea
\sum_{n=0}^{\infty} \frac{\Gamma(n-2\delta_\gamma)
\Gamma(n-1-3\delta_\gamma)}{n!\,\Gamma(n+1/2-3\delta_\gamma)} 
= \frac{\Gamma(-2\delta_\gamma)\Gamma(-1-3\delta_\gamma)
\Gamma(3/2+2\delta_\gamma)}{\Gamma(1/2-\delta_\gamma)\Gamma(3/2)} 
= -\frac{1+4\delta_\gamma}{6\delta^2_\gamma(1+3\delta_\gamma)} \,
\frac{\Gamma(1-2\delta_\gamma)\Gamma(1-3\delta_\gamma)
\Gamma(1/2+2\delta_\gamma)}{\Gamma(1/2-\delta_\gamma)\Gamma(1/2)}
\, .
\label{sum1}
\eea
The second term in the r.h.s. of Eq.~(\ref{split}) comes with a factor of
$2\delta_\gamma$. It is singular and, similarly to (\ref{term2}), the singular
part is only in the $n=0$ term of the series. So, we have
\be
\sum_{n=0}^{\infty} \frac{\Gamma(n-2\delta_\gamma)
\Gamma(n-3\delta_\gamma)}{n!\,\Gamma(n+1/2-3\delta_\gamma)}\,
\frac{1}{n-1/2-2\delta_\gamma} = 
\frac{\Gamma(1-2\delta_\gamma)
\Gamma(1-3\delta_\gamma)}{\Gamma(1/2-3\delta_\gamma)}\, 
\bigg[ -\frac{1}{3\delta^2_\gamma} + \frac{4}{3\delta_\gamma} -\frac{16}{3}
+ 16 G - 6\zeta_2 + O(\delta_\gamma) \bigg]\, .
\label{sum2}
\ee
With the help of Eqs.~(\ref{sum1}) and (\ref{sum2}), we obtain
%
\bea
\sum_{n=0}^{\infty} \frac{\Gamma(n-2\delta_\gamma)\Gamma(n-1-3\delta_\gamma)}{n!\,\Gamma(n-1/2-3\delta_\gamma)}\,\frac{1}{n-1/2-2\delta_\gamma} 
&&= \frac{\Gamma(1-2\delta_\gamma)\Gamma(1-3\delta_\gamma)}{\Gamma(1/2-3\delta_\gamma)}\, \bigg \{
-\frac{1}{6 \delta_\gamma^2} - \frac{1}{2\delta_\gamma } + \frac{25}{6} - 3\zeta_2 
\nonumber \\
&& \quad + \delta_\gamma \, \Bigl[-\frac{41}{2} + 32 G - 9\zeta_2 + 7\zeta_3 \Bigr] + O(\delta^2_\gamma) \bigg \}\, .
\eea
Combining all terms up to $O(\delta_\gamma)$ yields:
\bea
&&I(1/2) = \frac{\Gamma(1-2\delta_\gamma)}{24 \sqrt{\pi}}\,\bigg \{ -\frac{1}{\delta_\gamma^3} - \frac{7 -12 \ln 2}{\delta_\gamma^2} +
\frac{1}{\delta_\gamma}\,\Bigl( 61 + 6 \zeta_2 + 84 \ln 2 - 72 \ln^2 2 \Bigr) \bigg .
\nonum \\
&& \bigg . \quad -311 + 42\zeta_2 + 30\zeta_3 + 384\, G - 156 \ln 2 - 72 \,\zeta_2 \ln 2 - 504 \ln^2 2 + 288 \ln^3 2 + O(\delta_\gamma)  \bigg \}\, . 
\eea
Therefore:
\bea
&&G(1/2,1,1/2,1,1) = 
e^{-2\gamma_E \delta_\gamma}\,\bigg(\frac{1}{\delta_\gamma^2} + \frac{8(1-\ln 2)}{\delta_\gamma} - 56 - 5 \zeta_2 - 64 \ln 2 + 32 \ln^2 2  \bigg .
\nonum \\
&& \quad \bigg . + \delta_\gamma\, \Bigl( 240 - 40 \zeta_2 - \frac{110}{3}\,\zeta_3 - 384 \,G - 128 \ln 2 + 40\, \zeta_2 \ln 2 + 256 \ln^2 2 - \frac{256}{3} \ln^3 2 \Bigr) 
+ O(\delta_\gamma^2) \bigg)\, .
\label{G-RQED32}
\eea
%

Substituting the result of Eq.~(\ref{G-RQED32}) in Eq.~(\ref{sigma-c3}), yields:
\bea
&&e^{2(\gamma_E-L_p)\delta_\gamma}\,\sigma_{c_3}^{(2)} = \frac{\pi}{6\delta_\gamma^2} + \frac{\pi\,\big(3 - \bar{L}_p \big)}{3 \delta_\gamma}
+\pi\,\bigg( -11 - \frac{5}{2}\,\zeta_2  - 2 \bar{L}_p +  \frac{1}{3}\,\bar{L}_p^2 \bigg) 
\nonum \\
&& \quad + \pi \,\delta_\gamma\,\bigg( \frac{193}{3} - 5 \zeta_2 - \frac{55}{9} \zeta_3 - 64\,G - 96 \ln 2 + 22 \bar{L}_p + 
\frac{5}{3}\,\zeta_2\,\bar{L}_p + 2 \bar{L}_p^2 - \frac{2}{9}\,\bar{L}_p^3 \bigg)
+ O(\delta_\gamma^2)\, .
\eea
All divergent terms cancel each-other in the crossed-photon diagram which therefore turns out to be IR finite:
\bea
e^{2(\gamma_E-L_p)\delta_\gamma}\,\sigma_{c}^{(2)} = -2 \pi (1-a)^2
+ 4\pi \,\delta_\gamma\,\Bigl( 12 + 4 a -16\,G + (1-a)^2 ( \bar{L}_p - 3 \ln 2) \Bigr) 
+ O(\delta_\gamma^2)\, .
\eea
The total two-loop self-energy then reduces to:
\be
e^{2(\gamma_E-L_p)\delta_\gamma}\,\sigma_2 = -4 \pi N_F - \pi(1-a)^2  
+ 2\pi \,\delta_\gamma\,\Bigl( 2N_F(1 + 2\bar{L}_p - 8 \ln 2) + 25 + 10 a -3a^2 -32\,G + (1-a)^2 \bar{L}_p \Bigr)
+ O(\delta_\gamma^2)\, .
\ee
The theory is therefore finite ($Z_\psi = 1$) and the expression of the dressed fermion propagator reads:
\bea
&&-i {\slashed p}\,S(p) = 1 + \frac{\tilde{\al}}{4\pi}\, \sqrt{\pi} \bigg( 1-a +\delta_\gamma \big(4 - (1-a)\bar{L}_p \big) + O(\delta_\gamma^2) \bigg) 
\nonum \\
&&\quad +  \left( \frac{\tilde{\al}}{4\pi} \right)^2\,\bigg(-4 \pi N_F + 
4\pi \,\delta_\gamma\,\Bigl( N_F(1 + 2 \bar{L}_p - 8 \ln 2 ) + 16(1-G)-\frac{3}{2}\,(1-a)^2  \Bigr) + O(\delta_\gamma^2) \bigg) + O(\tilde{\al}^3)\, .
\eea
Remarkably, the $O(1)$ two-loop correction is gauge-invariant and reduces to a very simple form: $-4\pi N_F$ while the $O(\delta_\gamma)$
correction involves $\pi$, $\ln 2$ as well as the Clausen function $\text{Cl}_2(\pi/2)=G$.

\acknowledgments

I thank Anatoly Kotikov for discussions and help with the evaluation of the third part of
the integral $I(1/2)$.

\end{widetext}

\end{fmffile}
\end{document}